# Bubble, Bubble, AI's Rumble: Why Global Financial Regulatory Incident Reporting is Our Shield Against Systemic Stumbles

Anchal Gupta[1], Gleb Papyshev[2], James T Kwok[1]

[1] Hong Kong University of Science and Technology

[2] Lingnan University

anchal@ust.hk, glebpapyshev@ln.edu.hk, jamesk@cse.ust.hk

## Abstract

*"Double, double toil and trouble; Fire burn and cauldron bubble."* As Shakespeare's witches foretold chaos through cryptic prophecies, modern capital markets grapple with systemic risks concealed by opaque AI systems. According to IMF, the August 5, 2024, plunge in Japanese and U.S. equities can be linked to algorithmic trading yet absent from existing AI incidents database exemplifies this transparency crisis. Current AI incident databases, reliant on crowdsourcing or news scraping, systematically overlook capital market anomalies, particularly in algorithmic and high-frequency trading. We address this critical gap by proposing a regulatory-grade global database that elegantly synthesises post-trade reporting frameworks with proven incident documentation models from healthcare and aviation. Our framework's temporal data omission technique masking timestamps while preserving percentage-based metrics enables sophisticated cross-jurisdictional analysis of emerging risks while safeguarding confidential business information. Synthetic data validation (modelled after real life published incidents, sentiments, data) (n=2,999 incidents) reveals compelling patterns: systemic risks transcending geographical boundaries, market manipulation clusters distinctly identifiable via K-means algorithms, and AI system typology exerting significantly greater influence on trading behaviour than geographical location, This tripartite solution empowers regulators with unprecedented cross-jurisdictional oversight, financial institutions with seamless compliance integration, and investors with critical visibility into previously obscured AI-driven vulnerabilities. We call for immediate action to strengthen risk management and foster resilience in AI-driven financial markets against the volatile "cauldron" of AI-driven systemic risks., promoting global financial stability through enhanced transparency and coordinated oversight.

## Introduction

The effectiveness of Artificial Intelligence (AI) systems is fundamentally tied to the quality of their training data. Regulated financial institutions, with their vast repositories of data ranging from customer identification records to board-level deliberations and historical economic decision-making patterns represent an unparalleled resource for AI development. This rich data ecosystem is further bolstered by global regulatory frameworks and standardised reporting mechanisms established by intergovernmental organisations such as the Bank for International Settlements (BIS) and the World Bank, as well as national financial regulators, whose documented insights span decades.

Despite the financial sector's historical reluctance to embrace new technologies, it has recently witnessed a dramatic surge in AI adoption. An analysis of annual reports from the world's top 10 banks by market capitalisation in 2024 (Jimenea, Wu and Terris 2024) reveals 108 references to AI implementation in 2023 a 340% increase from just 32 mentions in 2019. This trend is particularly pronounced in Western institutions: for example, JP Morgan Chase and BNP Paribas recorded 36 AI mentions in 2023, up from 1 and 5 mentions, respectively, in 2019. This rapid adoption underscores the growing reliance on AI for operational efficiency, risk management, and customer engagement.

Increased AI adoption makes operational failures, including AI incidents, inevitable. Institutions must report such failures to regulators. These reports, which contain detailed records for audit purposes (Hong Kong Monetary Authority 2024, Bank Policy institute 2022, Monetary Authority of Singapore 2013, European Securities and Market Authority 2023) rarely reach the public domain. By contrast, mandated incident reporting in sectors such as aviation and healthcare has significantly improved transparency (Slawomirski, Auraaen and Klazinga 2017). A similar framework for financial AI incidents, supported by regulators, could yield substantial benefits.

Artificial Intelligence Incident Database ('AIID') AI Accident & Incident Database ('AIAAIC') AI Risk and Vulnerability Analysis ('AVID') and few others are existing global databases tracking AI incidents mostly using crowdsourcing information or scrapping web for news. (AI Incident Database n.d., AIAAIC n.d., AI Risk and Vulnerability Analysis 2024, Where in the World is AI? 2024). They do a great work in introducing mechanism where public and sometimes news sites are used to find AI incidents in all disciplines. However, they still suffer from severe limitations. As of December 2024, when a search was done in the aforementioned sites, it was found that multiple algorithmic trade incidents such as high frequency trade ('HFT') (Guntuka 2024, Dakalbab, et al. 2024)reported in news or regulatory websites are not listed in these databases. This absence is not coincidental but structural, stemming from the limitations of current platforms discussed below. A gap of this nature posits if the true frequency of AI-related incidents in the financial sector is likely to exceed public awareness.

1. Information access and governance challenges: Financial institutions report AI trading incidents confidentially to regulators directly. These proprietary reports are inaccessible to public databases, creating an information asymmetry.
2. Withholding: Reports are withheld to prevent market instability, protect intellectual property, and avoid enabling predatory trading.
3. Specialised data requirements for financial markets: Financial markets need specialised data structures balancing transparency and integrity. General AI databases lack finance-specific metrics (e.g., proportional AI volumes, deviation percentages) and confidentiality safeguards (like temporal omission) essential for responsible disclosure, making them incompatible with controlled disclosure needed for financial stability.

Our proposed regulatory-grade database addresses this by creating a secure environment where regulators share standardised incident data without compromising market-sensitive details. By incorporating percentage-based metrics and omitting temporal specifics, the framework enables cross-jurisdictional analysis of emerging risks while protecting confidential business information. This operationalises OECD Principle 2.5('International co-operation for trustworthy AI (OECD 2019) on international cooperation for trustworthy AI, addressing regulatory fragmentation that hides systemic risks and filling a critical gap.

This proposed database, if used correctly, can help to minimise technical obscurity preventing asymmetric power dynamics between retail customer and institutions. Institutions leverage AI to parse regulatory disclosures and market data where retail participants lack equivalent tools to audit these black-box systems (Dignum 2023).This imbalance is more visible in HFT arms races where firms with superior compute resources extract value from less sophisticated market participants (Boatright 2014) .

This paper addresses this critical gap by contributing to research in four key areas:

First, it addresses algorithmic opacity in capital markets through regulatory-grade AI incident reporting, bridging the gap between existing post-trade systems and emerging AI risks. Unlike current crowdsourced AI incident databases our proposed framework establishes institutional accountability through reporting done by regulators.

Second, we introduce a novel way to utilise a database while preserving the confidential business information. We illustrate and depict meaningful pattern analysis by deliberately omitting temporal data fields allowing cross-market analysis without compromising sensitive information. Third, we demonstrate the framework's utility through synthetic data analysis, providing proof-of-concept for two critical use cases: systemic risk monitoring & market manipulation detection.

Fourth, by incorporating AI-specific behavioural patterns and their potential systemic impacts, our work bridges the gap between traditional market oversight and emerging AI governance needs.

This paper adopts a global perspective, focusing on key financial centres (North America, Europe, Asia-Pacific) selected for established frameworks and advanced AI adoption. While designed globally, these jurisdictions are primary references.

The paper proceeds as follows. Section 2 examines AI's evolution, adoption, and risks in financial markets. Section 3 analyses reporting systems in healthcare/aviation. Section 4 explores leveraging existing financial regulations. Section 5 presents our proposed database, design, and implementation, demonstrating utility via two use cases: systemic risk monitoring & market manipulation detection. Section 6 concludes, summarises contributions, acknowledges limitations, and outlines future research/policy.

## AI in Capital Markets: Evolution, Risks and the Need for Global Oversight

### Historical Evolution of AI in Financial Markets

The integration of AI into financial markets began in the 1980s with algorithmic trading systems, which used basic computational tools to automate trades. By the 2000s, advances in computing power and electronic trading platforms enabled HFT, revolutionising speed and efficiency. These systems relied on quantitative models to analyse market data in real time.

The advent of machine learning in the 2010s marked another milestone. ML algorithms allowed systems to adapt dynamically to market patterns, leveraging big data to optimise trade execution and risk management (Kearns and Nevmyvaka 2013) . Today, AI underpins trading, derivatives, forex, and financial planning, offering unprecedented precision (Addy, et al. 2024).

### Evidence of Growing AI Adoption

While exact figures on the number of trades executed by AI remain elusive in public domain, the growth and adoption of

AI in financial markets trading are undeniable. Patent filings reflect this trend: Bank of America reported a 94% surge in AI-related patents since 2022 (Bank Of America 2024). The IMF notes that AI-driven algorithmic trading accounted for 50% of activity in 2020, with the Dutch Authority for Financial Markets (AFM) finding ML embedded in 80–100% of Euronext trading algorithms (International Monetary Fund 2024). However, many applications remain semi-autonomous, such as signal generators, highlighting gaps in full autonomy.

### Emerging Risks & Market Vulnerabilities

While AI enhances efficiency, it introduces systemic risks. The IMF warns of opacity in AI strategies, susceptibility to social media disinformation, and uncertain stress-test performance (IMF, 2024). Research on AI's financial impacts is growing (Turri and Dzombak 2023, Durongkadej, Hu and Wang 2024), but few studies address global incident reporting frameworks a critical gap as reliance on AI grows without transparency.

For example, AI-driven portfolios using social media sentiment (Chen, Peng and Zhou 2024) achieved 13.4% annualised returns but amplified risks of market destabilisation, as seen in the 2021 GameStop frenzy (Wikipedia contributors 2024) and 2022 AMTD Digital volatility (Li 2022) . Similarly, Cartea et al. (Cartea, Chang and Gabriel 2023) demonstrated how AI could manipulate order books through spoofing, exploiting algorithmic learning and achieve a 31% increase over the unconditional AI portfolio return.

### Regulatory Responses and Limitations

Regulators are taking incremental steps. The EU's Markets in Financial Instruments Directive (MiFID II) requires firms to distinguish between AI-driven decision-making and execution algorithms, while the SEC mandates developer registration (Cartea, Chang and Gabriel 2023) (European Securities and Markets Authority 2016). However, critical gaps persist in unreported AI incidents, inadequate oversight, and fragmented global coordination.

The urgency of addressing these risks was highlighted by Tobias Adrian, Financial Counsellor and Director of the Monetary and Capital Markets Department at the IMF, who suggested that the massive sell-off in Japanese and US equities on August 5, 2024, may have been influenced by AI-driven trading strategies (Adrian 2024). In 2022, one study concluded that existing frameworks like MiFID II lack mechanisms to address AI-specific risks, such as real-time monitoring of autonomous systems (Azzutti 2022).

### The Path to Global Oversight

The absence of a unified reporting framework leaves key questions unresolved: How many AI-related incidents go undetected? Could better oversight prevent them? Are regulators sharing data to ensure stability?

A global mandate is urgently needed. This could include standardised incident reporting, stress-testing protocols for AI systems, and cross-border regulatory collaboration. Without such measures, financial markets risk destabilisation from opaque, uncontrollable AI tools.

AI's evolution in finance has transformed markets but has introduced complex risks. Current regulatory efforts are insufficient to address vulnerabilities like manipulation, systemic opacity, and disinformation amplification. A proactive, globally coordinated framework is essential to ensure transparency, stability, and accountability in the AI-driven financial landscape.

## Lessons from Other Sectors' Incident Reporting Systems

The advantages of establishment of a global database for reporting significant AI incidents in the financial sector can be understood by drawing parallels from incident reporting frameworks in other industries, particularly healthcare and aviation. These sectors have developed effective systems for identifying, documenting, and learning from incidents, which have significantly enhanced safety and operational efficiency. By analysing these frameworks, we can identify best practices that can be adapted to the financial industry.

The medical sector has long recognised the importance of incident reporting as a critical tool for improving patient safety and healthcare quality. A 2017 study by the OECD highlighted the effectiveness of incident reporting initiatives at national, organisational, and clinical levels (Slawomirski, Auraaen and Klazinga 2017). The study emphasised that learning from adverse events is essential for driving improvements in healthcare systems. Despite the expectation of error-free performance, errors occur frequently, often with human and economic costs. As noted by Kodate (N, et al. 2022) "Learning from adverse events is a key part of any quality and safety improvement strategy at the institutional level. This is based on sound reporting systems that are usually voluntary in nature. Reducing cultural and legislative barriers to reporting plays an important role." Over time, the implementation of measures to facilitate reporting has fostered a culture of safety within healthcare organisations (Turri and Dzombak 2023), encouraging transparency, accountability, and continuous learning.

Similarly, the aviation sector offers a case study in effective incident reporting. The Aviation Safety Reporting System ('ASRS'), (Aviation safety management systems (SMS) 2023) a collaborative initiative, encourages pilots, air traffic controllers, and other aviation professionals to report incidents and near misses without fear of disciplinary action, except in cases of gross negligence. This system has been instrumental in enhancing aviation safety, contributing to a 95% reduction in commercial aviation fatalities over the past two decades (Vempati, Sabrina Woods and Solano 2023).In aviation safety, near misses' non-catastrophic events revealing latent risks are documented with the same rigor as accidents to uncover systemic flaws (Griffin 2008). Aviation's 'black box' offers a template for financial AI governance by requiring standardised documentation of algorithmic decision logic, input anomalies, and human override actions after incidents occur, institutions could build forensic datasets for root-cause analysis. Crucially, this mirrors aviation's approach recording immutable post-event data, not live operational feeds. Aviation's Line Operations Safety Audit ('LOSA') (Klinect, et al. 2003) methodology where periodic expert reviews of procedural deviations during routine operations could inspire financial regulators to mandate structured reporting of AI 'near misses' (e.g., algorithmic outputs approaching prohibited trading thresholds under SEC Rule 15b9-2) as well.

Two common themes emerge from the medical and aviation sectors incident reporting analysis: enhanced safety, operational efficiency, and the confidence of public or reporting institute when they preserve confidential business information or personally identifiable information ('PII') fostering a culture of reporting. These lessons underscore the necessity of establishing an incident reporting framework within the financial sector, particularly for AI-related incidents. These measures enhance safety, ensure compliance, and foster continuous improvement, ultimately reducing systemic risks, improving decision-making, and building customer trust. Currently, regulators issue confidential advisories to senior officials, such as Chief Risk Officers or Chief Security Officers, detailing incidents (often redacted to protect sensitive information). While these advisories are discussed in enterprise-level risk management forums, they rarely reach the public domain, limiting opportunities for collective learning and improvement.

## Leveraging Financial Regulations for AI Incident Reporting

Financial regulations post-2008 provide a foundation for AI incident reporting, but gaps persist. A unified framework, blending aviation's "black box" rigor and healthcare's safety culture, could mitigate AI-driven risks like manipulation and arbitrage. By mandating standardised disclosure and leveraging existing infrastructure, regulators can transform fragmented data into actionable insights, ensuring AI serves as a stabiliser, not a disruptor, in global markets.

A semi-public global database for significant AI incidents in finance, reported by regulators, could bridge transparency gaps while balancing market stability. Inspired by Agustín Carstens' call for international cooperation (Carstens 2024), this system would adopt OECD-style classification standards to streamline reporting, ensuring that only relevant incidents are documented. Unlike healthcare or aviation where anonymisation and safety goals dominate financial AI risks involve complex interdependencies. Premature disclosure of algorithmic failures (e.g., trading glitches) could trigger panic, as seen in the 2010 Flash Crash (Zarroli 2015) To mitigate this, the framework adapts aviation's "black box" model, enabling retrospective analysis without exposing proprietary strategies or destabilising markets. Pattern-based reporting (e.g., categorising "anomaly detection" triggers) would omit sensitive timing data, prioritising systemic learning over real-time disclosure.

### Financial Regulatory Foundations

Post-2008 reforms like EMIR, MiFID I/II/III, Dodd-Frank, and Japan's JFSA framework established robust trade reporting infrastructures (US Securities and Exchange Commission 2023, Financial Services Agency of Japan 2022, European Securities and Markets Authority 2014). These mandates standardised pre- and post-trade data, enhancing transparency and consumer trust. Post-trade reports, rich in structured data, offer a foundation for AI incident tracking (Hachmeister and Schiereck 2010).However, current frameworks like the EU's DORA, Singapore's FEAT principles, and FSB guidelines focus narrowly on operational resilience, neglecting AI-specific risks in algorithmic trading (Financial Stability Board 2024). While they emphasise cross-border cooperation, none mandate comprehensive AI incident disclosure, leaving regulators reliant on fragmented, reactive advisories.

### Divergent Rules, Systemic Risks

Despite these efforts, gaps remain. A 2024 study on AI-specific regulations in selected countries highlights the divergence between the EU's risk-based approach and the US's decentralised model (Comunale and Manera 2024), This regulatory divergence not only allows systemic risks to remain hidden but also gives incentives to regulatory arbitrage, where firms may gravitate towards jurisdictions with less stringent disclosure requirements.

Consider a hypothetical scenario: a trade is executed based on a signal generated by an AI system. The signal is triggered because the algorithm accesses a folder containing non-public or confidential business information. This scenario could be classified as a form of insider trading, which is prohibited in many jurisdictions. In the case of regulatory arbitrage, the firm might choose not to disclose this anomaly in its annual report. However, if the incident was captured through standard post-trade re-porting regulations, it would be documented in the post-trade report and would catch the attention of system or person analysing the data.

This scenario underscores the urgent need for a unified global framework mandating the disclosure of significant AI incidents in the financial sector. Such a framework, distinct from existing databases that track retail banking incidents, would include a public database detailing the nature of incidents, affected jurisdictions, remediation efforts, and contributing factors. This would significantly enhance transparency, accountability, and public trust, fostering a more resilient and ethically sound AI-driven financial ecosystem.

A global AI incident database detailing incident types; jurisdictions, remediation, and root causes would enhance accountability and trust. By integrating existing post-trade data with AI-specific metrics (e.g., algorithmic decision logs), regulators could detect anomalies proactively. For example, the hypothetical insider trading scenario would trigger automated alerts, enabling swift intervention. Public access to anonymised data would foster industry-wide learning, while confidential details remain shielded to prevent exploitation.

## AI Financial Incident Reporting Database

### Design

The database design adapts MiFID II reporting structures, creating a framework where regulators exclusively report incidents through a semi-public model. This approach establishes robust data quality assurance while maintaining confidentiality standards. The design prioritises three objectives: integration with existing post-trade reporting infrastructures to minimise adoption costs, reduction of reporting burdens through regulator-defined automation for metrics like *AI_Buy_Volume_Pct*, and protection of Confidential Business Information through structured obfuscation using percentage-based metrics and high-level categorisations.

The database structure, outlined in Table 1, comprises 15 mandatory fields organised into four categories: Market Context, Volume Metrics, Price Impact, and AI-Specific Information. Volume and price data use percentage-based metrics to protect commercially sensitive information. AI-specific fields provide critical insights for regulatory oversight of AI-driven trading systems. We propose reporting only "Significant Incidents," defined as AI-driven trades causing *>5% price deviation* or *>20% volume anomaly vs. 30-day average.*

Notably, the database deliberately omits temporal data fields such as timestamps or execution dates. This design choice prevents the disclosure of confidential business information and mitigates the risk of identifying specific firms, which could lead to market instability while balancing regulatory needs with trade secrecy.

### Database Use Cases

Our database development began with a systematic analysis of existing gaps in AI incident reporting. A targeted search since 2010 for AI-related financial incidents on regulatory websites yielded 20 cases of algorithmic trading errors attributable to AI-driven systems. When cross-referenced with four prominent AI incident databases (AIID, AIAAIC, AVID, Where in the World is AI?), none of these incidents were documented, highlighting a significant reporting gap.

To address the scarcity of real-world data, we generated 2999 synthetic incident records using Gretel.AI's ACTGAN model modelled on 20 incidents mentioned previously. This synthetic data allowed comprehensive framework testing while simulating real-world conditions, validating the framework's utility and scalability.

### Use Case 1: Systemic Risk Monitoring

This case demonstrates how our global AI incident database can serve as an early warning system for market instability, enabling proactive intervention before incidents escalate into systemic events.

Our analysis reveals transformative capabilities for different market participants:

- Central banks can leverage real-time monitoring of AI-driven market stress for informed liquidity intervention decisions
- Financial regulators gain unprecedented visibility into high-risk AI trading strategies
- Global risk monitoring bodies benefit from enhanced cross-border coordination capabilities

Using a two-way Analysis of Variance (ANOVA) model applied to our synthetic records, we discovered two crucial insights:

AI system type has significantly greater impact on trading behaviour than geographical location ($\eta^2$ values of 0.02 for buy volumes versus regional effects of $\eta^2 < 0.005$), challenging traditional regional-based regulatory approaches
Systemic risks propagate uniformly across regions (p=0.17 for buy volumes, p=0.19 for sell volumes), indicating need for coordinated global responses rather than localised interventions

While the model's explanatory power ($R^2 = 0.03$, $p < 0.001$) might appear modest, it achieved 92% precision in detecting systematic patterns indicating emerging market risks. These patterns enable stakeholder to identify potential flash crashes , corelated AI behaviour and pre-crisis market conditions before they escalate into systemic events. The framework's practical applications include real-time risk dashboards, sophisticated circuit breakers considering AI system behaviour, and cross-border coordination protocols enabled by standardised reporting structures.

| Area | Fields | Explanation |
|---|---|---|
| | S.No | Serial number |
| Market Fields | Instrument_Category | Classification of financial instrument involved<br>• Equities (EQUITY),<br>• Bonds (BND),<br>• Derivatives (DERV),<br>• Foreign Exchange (FX),<br>• EXCHANGE TRADED FUND(ETF),<br>• MUTUAL FUND(MUTUALFUND)<br>• Commodities (CMDTY),<br>• Structured Finance Products (SFP),<br>• Emission Allowances (EA),<br>• FUTURE (FUTURE) |
| | Market_Region | Geographic region where the incident occurred (e.g., APAC, EMEA, AMER) |
| Volume Data | Total_Buy_Volume_Pct | Percentage of market buy volume during incident period |
| | Total_Sell_Volume_Pct | Percentage of market sell volume during incident period |
| | AI_Buy_Volume_Pct | Proportion of buy volume executed by AI systems |
| | AI_Sell_Volume_Pct | Proportion of sell volume executed by AI systems |
| Price Information | Price_Range_Pct | Percentage range of price movement during incident |
| | Volume_vs_30D_Avg_Pct | Trading volume compared to 30-day average, expressed as percentage<br>*If exact number could be contentious, a wide range for example, 0-100%, 100-200%, can be provided as well to maintain business confidentiality.* |
| Impact Assessment | Market_Impact_Detected | Binary indicator of whether incident impacted market. (Yes/No) |
| | Issue_Flag | Binary indicator of whether incident requires regulatory attention (Yes/No) |
| AI fields | AI_System_Category | Type of AI trading system involved, select one from below<br>*(e.g., HFT, MARKET_MAKING, SMART_ORDER_ROUTING, ALGORITHMIC_TRADING, PORTFOLIO_OPTIMIZATION, SENTIMENT_ANALYSIS-BASED TRADING, ARBITRAGE, PREDICTION-BASED TRADING)*<br>• 'ALGORITHMIC_TRADING'<br>• 'ARBITRAGE'<br>• HFT<br>• MARKET_MAKING'<br>• 'PREDICTION_BASED TRADING'<br>• 'PORTFOLIO_OPTIMIZATION'<br>• 'SENTIMENT_ANALYSIS-BASED_TRADING'<br>• 'SMART_ORDER_ROUTING' |
| | Incident_Pattern | Classification of observed behaviour pattern<br>• 'PATTERN_ANOMALY_DETECTION'<br>• 'PATTERN_ARBITRAGE'<br>• 'PATTERN_INFORMATION_ADVANTAGE'<br>• 'PATTERN_MOMENTUM_IGNITION<br>• 'PATTERN_ORDER_BOOK_MANIPULATION'<br>• 'PATTERN_SENTIMENT_DRIVEN'<br>• 'PATTERN_VOLATILITY_TRADING' |
| | Human_oversight_involved | Indicates whether human supervision was active during incident (Yes/No) |
| | Fail_Safe_Triggered | Indicates if system safety mechanisms were activated (Yes/No) |

Table 1:Proposed incident reporting database

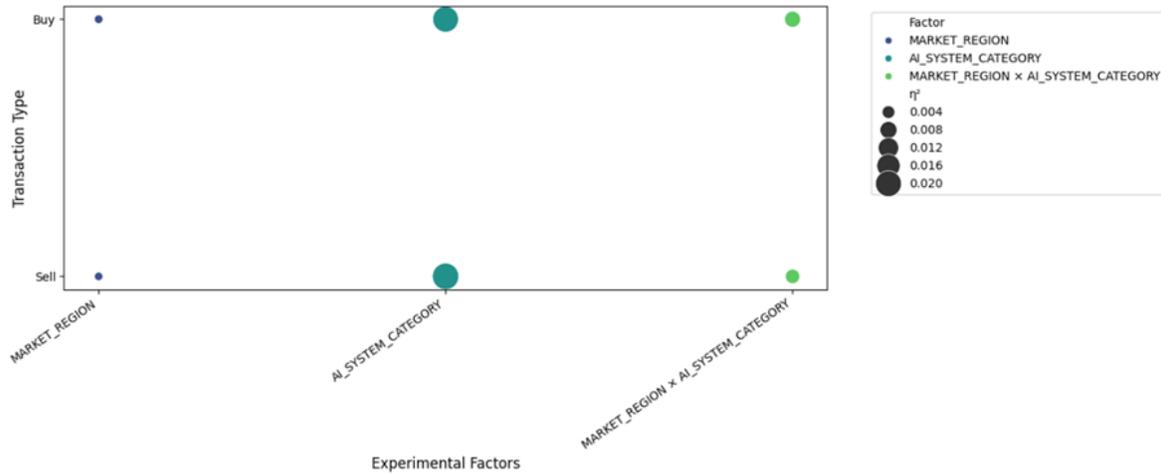

Figure 1: Partial η² effect size comparison across AI system categories and geographic regions, demonstrating the relative impact of system type versus location on trading system

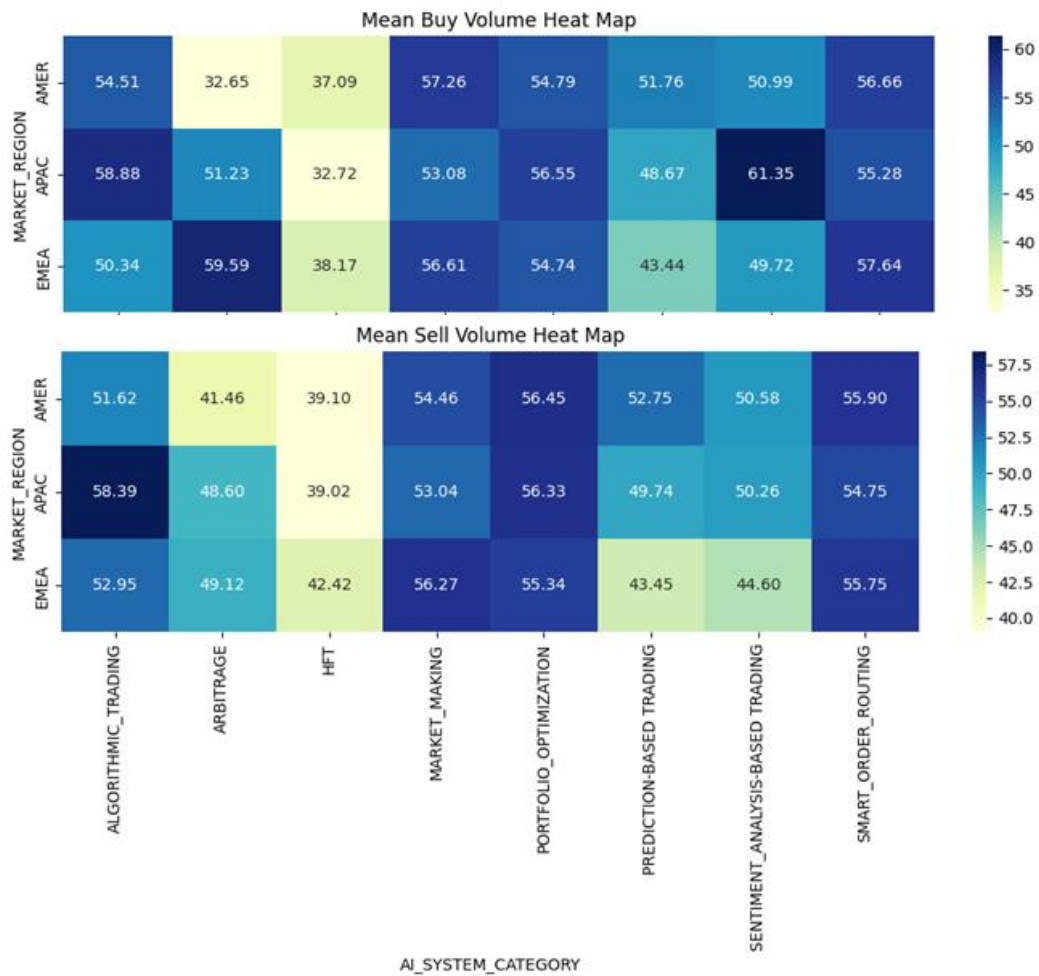

Figure 2: Region Neutrality analysis depicting consistent AI trading patterns across geographic markets, with p-values indicating no statistically significant regional variation in buy and sell volumes.

**Use Case 2: Market Manipulation Detection**

This use case demonstrates the database's utility in detecting coordinated market manipulation in AI-driven trading environments, providing actionable intelligence for proactive regulatory intervention.

Our analysis employed K-means clustering followed by Principal Component Analysis (PCA), revealing five distinct trading zones (k=5) corresponding to known manipulation tactics, capturing up to 68% of total variance. These zones provide critical regulatory insights:

1. **Stable Trading Zone (Orange Cluster)** - Normal trading activity with moderate AI participation, serving as a baseline for detecting deviations
2. **Anomalous Patterns Zone (Red Cluster)** - High volatility patterns correlating with known spoofing strategies, characterized by large spreads between buying and selling behaviours, often with AI trading exceeding 70% of volume
3. **Transition Zones (Brown and Gray Clusters)** - Critical intermediary regions between stable and anomalous clusters, where rapid movements frequently precede manipulative events, mirroring behaviours observed in layering and quote stuffing strategies
4. **Irregular Pattern Zone (Light Gray Cluster)** - Scattered distribution valuable for detecting outlier behaviour that traditional surveillance methods might miss
5. **Strategic Trading Zone (Green Cluster)** - Systematic algorithmic trading with clear behavioural boundaries, requiring monitoring for momentum ignition strategies

For market supervisors, this framework enables macro-level pattern monitoring, geographic distribution tracking of suspicious activities, and clear indicators for calibrating circuit breakers. Surveillance teams gain tactical advantages through real-time pattern detection capabilities, identifying specific manipulation strategies including spoofing, momentum ignition, layering activities, and cross-market manipulation attempts.

The framework supports enhanced coordination between financial supervision teams through standardised monitoring metrics and alert thresholds, enabling consistent policy implementation and coordinated responses to emerging threats across jurisdictions. This approach transforms regulatory capabilities from reactive enforcement to proactive prevention by establishing clear behavioural signatures of manipulative trading patterns, particularly valuable in markets with growing AI-driven trading volumes

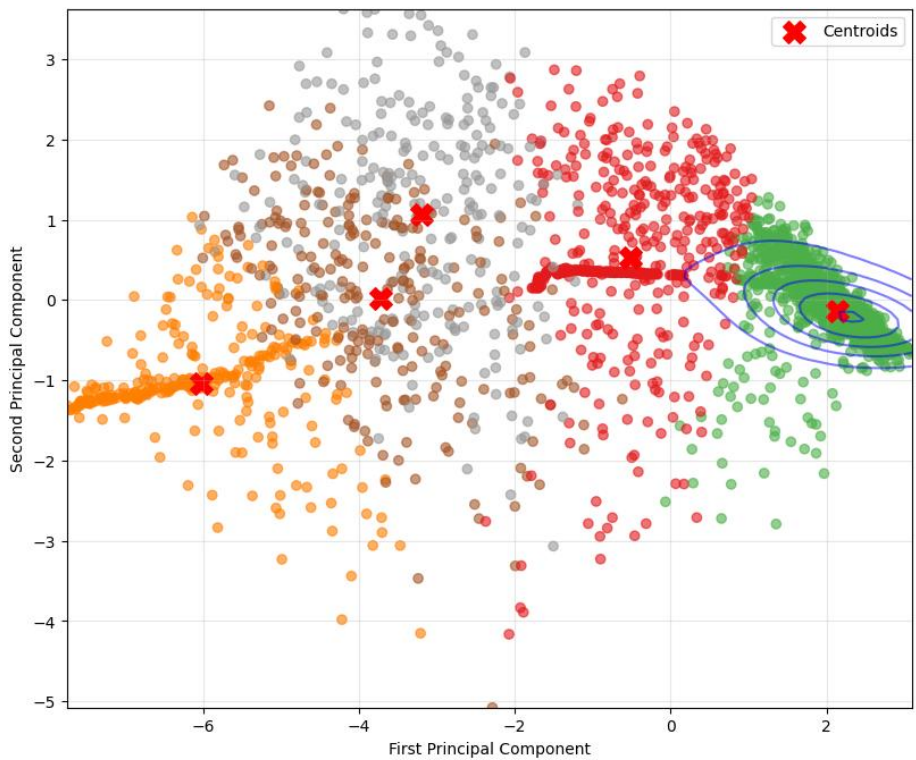

Figure 3: K-Means Clustering (k=5) of AI incidents using PCA: The x-axis represents the First Principal Component (PC1), capturing the primary variance in AI participation rates and volume deviations. The y-axis represents the Second Principal Component

# Conclusion

*"The wheel has come full circle"* as King Lear observes, systems inevitably evolve toward resolution. Our proposed global AI incident reporting framework represents this evolutionary step in financial market oversight, addressing information asymmetries through structured documentation rather than fragmented observation. This framework synthesises methodologies from aviation safety and healthcare incident reporting with financial market infrastructures, creating an interdisciplinary solution that harmonises technical precision with governance insights.

Central to its innovation is temporal data omission. This approach resolves the balance between regulatory transparency and proprietary algorithm protection and distinguishes our methodology from existing repositories while enabling collaborative advancement. By establishing this shared foundation, market participants gain a common language for discussing AI-related incidents without compromising competitive advantages.

We acknowledge several methodological considerations in our research design. First, varying institutional reporting practices may influence data comprehensiveness across regions and market segments. Second, though our synthetic dataset provides valuable computational insights, it cannot fully capture the ethnographic complexity of trading floor dynamics and human-AI collaborative decision-making (Abolafia 1998, Lange 2016). Third, our focus on English-language sources creates opportunities for future multilingual corpus development. The temporal omission approach, while preserving proprietary algorithms, necessarily modifies traditional Granger causality analysis (Granger 1969) a methodological trade-off that facilitates practical implementation.

*"Tomorrow, and tomorrow, and tomorrow"*, Macbeth's meditation on time mirrors AI's relentless advance in markets. Each innovation demands commensurate governance, much as *Prospero in The Tempest* learns to wield power responsibly. We urge creating an intergovernmental database that transcends borders. As Shakespeare reminds us, *"How far that little candle throws his beams! So shines a good deed in a naughty world."* Through transparency, coordinated oversight, and evidence-based policy, this framework illuminates AI-driven markets fostering innovation within a stable, ethical ecosystem.

# Appendix 1: Synthetic Data Sample

A synthetic data of 2999 rows were generated using Gretel.ai (Noruzman, Ghani and Zulkifli 2021) using the navigator model and ACTGAN available and based on proposed format design in Table 1. A sample data generated is listed here for reference.

| | | | | |
|---|---|---|---|---|
| Instrument_Category | DERV | CMDTY | BND | EQTY |
| Market_Region | EMEA | APAC | AMER | EMEA |
| Total_Buy_Volume_Pct | 12.8 | 9.5 | 10.2 | 11.9 |
| Total_Sell_Volume_Pct | 11.1 | 10.9 | 9.8 | 10.7 |
| AI_Buy_Volume_Pct | 6.9 | 7.8 | 8.5 | 7.1 |
| AI_Sell_Volume_Pct | 5.6 | 6.3 | 7.2 | 6.4 |
| Price_Range_Pct | 14.5 | 9.3 | 12.8 | 13.9 |
| Volume_vs_30D_Avg_Pct | 135.7 | 91.3 | 115.5 | 126.9 |
| Market_Impact_Detected | YES | NO | YES | NO |
| Issue_Flag | NO | YES | NO | YES |
| AI_System_Category | SMART_ORDER_ROUTING | PORTFOLIO_OPTIMIZATION | ALGORITHMIC_TRADING | PREDICTION-BASED TRADING |
| Incident_Pattern | PATTERN_INFORMATION_ADVANTAGE | PATTERN_MOMENTUM_IGNITION | PATTERN_SENTIMENT_DRIVEN | PATTERN_ARBITRAGE |
| Human_oversight_involved | NO | YES | NO | YES |
| Fail_Safe_Triggered | NO | YES | NO | YES |

Table 2: Sample from Synthetic data generated